# Wetland Quality as a Determinant of
# Economic Value of Ecosystem Services: an Exploration

H Chen[a]*, P Kumar[b] and T Barker[a]

[a] *School of Environmental Sciences, Liverpool University, UK*
[b] *Ecosystem Services Economics Unit, UN Environment Programme*

*Email: h.chen47@lancaster.ac.uk; hongyan.chen88@gmail.com

## Abstract

Wetland quality is a critical factor in determining values of wetland goods and services. However, in many studies on wetland valuation, wetland quality has been ignored. While those studies might give useful information to the local people for decision-making, their lack of wetland quality information may lead to the difficulty in integrating wetland quality into a cross-studies research like meta-analysis. In a meta-analysis, a statistical regression function needs withdrawing from those individual studies for analysis and prediction. This research introduces the wetland quality factor, a critical but frequently missed factor, into a meta-analysis and simultaneously considers other influential factors, such as study method, socio-economic state and other wetland site characteristics, as well. Thus, a more accurate and valid meta-regression function is expected. Due to no obvious quality information in primary studies, we extract two kinds of wetland states from the study context as relative but globally consistent quality measurement to use in the analysis, as the first step to explore the effect of wetland quality on values of ecosystem services.

## 1. Introduction

Wetland ecosystems are one of the most productive ecosystems on this planet, delivering massive goods and services to human society (Maltby 2022). However, due to poor awareness of their values and underestimation of their contribution, many wetlands have been converted to farmland or urban areas or influenced by pollution due to agricultural and industrial activities (Maltby 1986). Consequentially, global wetland ecosystems have severely declined and degraded during the past decades. In order to restore and protect wetlands, hence ensure a sustainable supply of wetland goods and services, it is important to recognize their values. Vital to this is the development of valuation methods that explicitly link wetland values, the capital base of the ecosystem, to the design of policies (Pearce and Atkinson, 1993; Dasgupta and Mäler, 2000; Arrow *et al.*, 2004; Maler *et al.*, 2008; Dasgupta, 2010; Kumar and Chen, 2014).

For a typical wetland ecosystem, its values can be accounted in terms of populations and diversity of species, and annual rates ($a^{-1}$) of fish harvested, carbon stored or recreational visits. These are generally categorised as values from wetland provisioning, regulating or cultural services (MA, 2005). Proper and accurate estimation of these values enables comparative analyses of intervention practices and contributes to improvements of policy-making (Barbier, 1993; Barbier *et al*., 1997; Turner *et al.*, 2000). A critical factor in determining the values of wetlands is quality since a healthy and functioning wetland will provide rich ecosystem services (Zedler and Kercher, 2005; Maltby, 2009). This is a neglected aspect of studies.

The quantity of the wetland valuation practice has increased in relatively recent years. In the review by Heimlich *et al.* (1998), 33 studies over the last 26 years were listed and in Brander *et al.* (2009) there are more than 50 valuation studies for European inland wetlands. Based on primary studies of wetland valuation, several meta-analyses have been conducted to explore the commonalities through inter-study comparisons, to find the general relationship between wetland values and influential factors and to estimate the wetland values of non-valued areas. The examples include Brouwer *et al.* (1999), Woodward and Sui (2002), Brander *et al.* (2006), Brander *et al.* (2009), Ghermandi *et al.* (2010).

The wetland quality factor has been ignored in most primary studies and subsequently in the meta-analyses based on them. As mentioned in Brander *et al*. (2007), it is often the case that the provision of goods and services is indicated in a meta-analysis merely by binary variables, and that quality is not captured at all. This limitation may lead to generalisation errors and therefore to benefit transfer errors, which would probably lead to errors in policy making for the wetland sustainable development.

This paper tries to integrate a wetland quality factor into a meta-regression function, keeping other influential factors, such as study method, socio-economic state and other site characteristics, included as well. Due to no obvious quality information in primary studies, we extract two kinds of wetland states from the study context as relative but globally consistent quality measurement to use in the meta-analysis. Thus, a more accurate and valid regression result is expected.

## 2. Overview of wetland quality evaluation

Wetlands differ immensely with respect to their hydrology, geochemistry, plant communities and landscape position. In a real sense, no two wetlands are similar in their quality or function (Wright *et al.*, 2006). Despite this variability, some consistent and recurring impacts can be observed and some common dynamic processes and properties can be found in different wetlands. Wetland quality is to measure if the wetland is healthy or not, which might be reflected in the performance of wetland functioning. Wetland hydrological, geochemical and ecological functions and their performance are the direct or indirect results of wetland characteristics and dynamic processes.

Various methods have been developed for wetland quality evaluation in the past decades. In USA, Wetland evaluation technique (WET) (Adamus *et al.,* 1987) assigns a qualitative probability rating of high, moderate or low to each function, according to characterizing predictors and their relationship with wetland functions. However, it does not provide information on the degree of functional performance. A more advance method might be the hydrogeomorphic (HGM) approach (Smith, 1994; Brinson *et al*., 1998; Cole, 2006), which provides a procedure for assessing the capacity of a wetland to perform functions, based on interactions of the structural components of the ecosystem with surrounding landscape features. From these methods, Maltby (2009) developed a wetland functional assessment method to define and assess wetland quality for European wetlands, combining the probability of occurring and the level of the performance of wetland functions.

Quality index presents an alternative to the above methods. This method compares wetland functional capacity under existing conditions with that under the conditions where the highest, sustainable functional capacity occurs (Smith *et al*., 1995). The quality index may measure the functional capacity either across the suite of functions performed by a wetland (Smith and Theberge, 1987) or for a specific function from a specific perspective. An example of the former can be found in Lodge *et al*. (1995), where the indices are derived by totalling the weighted scores for each function and dividing by the maximum possible score. The instances of the latter include the index of biological integrity developed by Karr *et al*. (1986), floristic quality assessment index by Miller and Wardrop (2006) and wetland fish index by Seilheimer and Chow-Fraser (2006).

### 3. Meta analysis and wetland quality: a review

As a technique originally used in experimental medical treatment and psychotherapy (Glass, 1976), meta-analysis statistically analyzes the findings of empirical studies, helping to extract information from large masses of data in order to quantify a more comprehensive assessment (Brouwer *et al*., 1999). It enables researchers to explain differences in outcomes found in single studies on the basis of differences in underlying assumptions, standards of design and/or measurement (Ledoux *et al*., 2001). Since the beginning of the 1990s, meta-analysis has been playing an increasingly important role in environmental economics research (van den Bergh *et al.* 1997; Smith and Kaoru, 1990). In economic valuation of ecosystem services, meta-analysis has been used to explore the relationship between values and their influential factors. This general functional relationship could be used to predict with local information the ecosystem values for policy sites where conducting a primary research is unfeasible (function transfer).

Several articles, for example, Brander *et al.,* (2006) and Bateman and Jones, (2003), have argued that function transfers perform better than direct transferring the estimated values in primary studies to the policy sites in question. However, for the use of meta-analytic functions for benefit transfer, some important underlying assumptions (Rosenberger and Phipps, 2007) have to be borne in mind:

1) There exists a valuation function that links the values of a resource with the characteristics of sites, studies and socio-economic circumstances;

2) The difference between sites can be captured through a price vector;

3) Values are stable over time or vary in a systematic way, and

4) The sampled primary valuation studies provide 'correct' estimates of resource value.

Meta-analysis has been used to investigate and explain the variation in wetland ecosystem services valuation. While it has been widely recognized that wetland functions are the physical basis for providing wetland services, it seems that the difference in quality of wetland functioning has not been captured in the wetland economic valuation meta-analyses.

Brouwer *et al.* (1999) conducted a meta-analysis on wetland contingent valuation studies and tried to separate the use values and no-use values according to four different wetland functions, highlighting that the goods and services are provided because of the performed functions. The quality of the wetlands in Brouwer *et al.* (1999) has been considered from three aspects: wetland size, wetland type and wetland functioning. While the size can be measured as a continuous numeric and the type as differentiable groups, the wetland functioning can be only judged by a binary variable. No further difference of the performance level can be distinguished.

Woodward and Sui (2002) did another meta-analysis on wetland economic valuation. In their research, wetland services were captured using qualitative variables, binary variables (again), to indicate whether the service is reflected in the value. Their paper extends the variability associated with valuation methods and reduces the variability in services that can be considered, comparing the one by Brouwer *et al.* (1999). However, wetland quality was not explicitly considered. Woodward and Sui (2002) suggest that it would be highly speculative to use a single point from this extracted distribution in a benefits transfer exercise (Woodward and Sui, 2002).

Ghermandi *et al.* (2010) carried out a meta-analysis, considering the degree of the environmental pressure. Their analysis finds that the coefficient of the environmental pressure is negative, tending to indicate that a higher pressure of human activities on the wetland produces higher values. Their explanation is that human activities contribute to translate potential uses into values or that human interventions in a wetland often aim to improve the level of provision of specific wetland services. However, as we know, urbanization and agricultural activities in the immediate surroundings of the wetland frequently lead to the degradation of the wetland quality (Verhoeven and Setter 2010). Therefore, the degree of pressure taken into account in the meta-analysis in this paper has no monotonous and inverse proportional relationship with wetland quality.

Brander *et al.* (2009) performed a meta-analysis to scale up the wetland values to a European level. In the analysis, they introduced scarcity of wetland within a given area and discussed distance delay effect on the economic values. Wetland quality has been a discussion topic in their paper and the limitation of meta-analyses to capture differences in the quality of the services under consideration has been recognised. As they motioned, this limitation may translate into transfer errors, as the estimated transfer function cannot reflect important quality differences in characteristics across sites.

## 4. Meta analysis on economic valuation of wetland ecosystem services

In this meta-analysis, we explicitly elicit wetland quality as a critical influential factor and explore its effect on economic values of wetland services.

### 4.1 Primary studies and their properties

TEEB database is built under the Economics of Ecosystems and Biodiversity (TEEB) study (http://www.teebweb.org/). It collects information from hundreds of books, journals and reports on ecosystem valuation and records the values estimated for 18 biomes and 30 ecosystem services. Primary studies are selected from the database.

Querying the database with 'inland wetlands' as a criterion of biome, we gain 255 items. Then we drop studies that don't meet some standards suggested by Smith and Pattanayak (2002). The dropping process is as the following: after the data pre-processing by removing repeated items and the items which are not of the 'value per annum' type, we obtained a list of 149 data items from 43 articles. Since the values derived with a benefit transfer method are not strictly primary studies, we therefore deleted those items. Subsequently, the values that are not of a single service but a total economic value (TEV) or marked as 'various' had to be moved from the list for consistence. As a result, 70 data items from 27 articles remain in the analysis.

The cross tables based on the data are given in Tables 1 and 2. Table 1 displays the relationship between wetland services and types in terms of the number of study cases, while Table 2 shows the relationship between the services and the methods used for their valuation. The most studied services are food and raw materials, which are addressed mostly in the wetland types of *floodplains* and of *unspecified by the authors*; the favourite valuation method is the direct market pricing method, there being 41 data items out of the total 70.

Table 1. Cross table of ecosystem services and types of wetland

| Eco-Service | Ecosystem | | | | Grand Total |
|---|---|---|---|---|---|
| | Flood-plains | Peat-wetlands | Swamps / marshes | Wetlands [unspecified] | |
| Climate | | | 1 | | 1 |
| Extreme events | 1 | 2 | 3 | | 6 |
| Food | 8 | | 3 | 8 | 19 |
| Genepool | | 1 | 1 | 2 | 4 |
| Medical | 1 | | | 1 | 2 |
| Ornamental | | | 1 | | 1 |
| Raw materials | 6 | | 4 | 7 | 17 |
| Recreation | 1 | 1 | 1 | 2 | 5 |
| Soil fertility | | | | 1 | 1 |
| Waste | 2 | 2 | 2 | 2 | 8 |
| Water | 2 | | 1 | 2 | 5 |
| Water flows | 1 | | | | 1 |
| Grand Total | 22 | 6 | 17 | 25 | 70 |

Table 2. Principal services and goods provided by wetlands and valuation methods commonly used to estimate their value

| Eco-Service | Valuation Method | | | | | | | Grand Total |
|---|---|---|---|---|---|---|---|---|
| | Avoided Cost | Contin-gent Valuation | Direct market pricing | Factor Income / Production Function | Mitigation and Restoration Cost | Replace-ment Cost | Travel Cost | |
| Climate | 1 | | | | | | | 1 |
| Extreme events | 5 | | | | 1 | | | 6 |
| Food | | | 17 | 2 | | | | 19 |
| Genepool | | 3 | 1 | | | | | 4 |
| Medical | | | 2 | | | | | 2 |
| Ornamental | | | 1 | | | | | 1 |
| Raw materials | | | 17 | | | | | 17 |
| Recreation | | 1 | 2 | 1 | | | 1 | 5 |
| Soil fertility | | | | | | 1 | | 1 |
| Waste | 1 | 1 | | 1 | 1 | 4 | | 8 |
| Water | 1 | | 1 | | 1 | 2 | | 5 |
| Water flows | | | | 1 | | | | 1 |
| Grand Total | 8 | 5 | 41 | 5 | 3 | 7 | 1 | 70 |

## 4.2 Wetland quality perceived in the primary studies

There is no field in the database defined to describe the quality of wetlands. In order to explore the impact of wetland ecosystem quality on their value, we go back to the original literature to search for quality information.

There exists little information on the wetland quality of the study sites in the literature. It is not hard to understand the lack of linkage between the valuation and physical quality at an individual study scale, especially when a market price method is used. However, when making a generalisation using those individual studies, it is hard to establish a functional relationship between the values and the quality states of the wetlands when the quality of wetland sites is unknown. Furthermore, the relationship which does not capture the commonality among the primary studies can hardly give a proper prediction.

To make the work done, we scrutinize the original articles to try to find some clues about site quality. The critical clues include the objective of the valuation, valuation method, natural landscape and human activities within or surrounding the site.

Since the collected information from the articles is not enough to perform any kind of evaluation described in Section 2, we extract two categories of quality states, naturally functioning and degraded, for the meta-analysis. This is a relative quality classification definition, because wetlands that fall into the same category may have different productivity and thus different values. However, it is also a global consistent definition, as wetlands, no matter what geographic conditions they possess, can be always assigned a quality of one of the two categories.

Table 3 lists the original articles, part of the description used for drawing a quality state and as well the assigned quality codes to give an overview on the quality perception in this paper.

**Table 3.** Quality information on study sites

| Article | Quality of wetlands | Description |
|---|---|---|
| Acharaya and Barbier, 2000 | 1 (naturally performed) | This analysis has been conducted in the Madachi *fadama,* a regularly inundated area with good groundwater stocks. The groundwater recharge function is assumed to support dry season agricultural production dependent on groundwater abstraction for irrigation. |
| Adekola *et al.* ,2008 | 2 (degraded) | The size of the Ga-Mampa wetland was halved between 1996 and 2004. Using a direct market valuation technique, we estimated the economic value of the main provisioning services provided by the wetland. … current use exceeds sustainability levels. --=+? |
| Barbier, et al., 1991 | 2 | The floodplain has come under increasing pressure from drought and the upstream water development. These benefits are substantial, even when taking into account the unsustainability, … |
| Costanza, et al., 1997 | 1 | The valuation approach taken here assumes that there are no sharp thresholds, discontinuities or irreversibilities in the ecosystem response functions. Table 2 reports only the average values. |
| Department of Conservation, 2007 | 1 | In December 1989, a 5690 ha portion of Whangamarino Wetland became formally recognised under the United Nations convention on world wetland conservation. The Ramsar designation was inspired by the native species and ecosystem values. |
| Dubgaard, et al., 2002 | 1 | The valuation is used for the cost-benefit analysis of the Skjern river project, which purpose is to re-establish a large coherent nature conservation area. |
| Emerton, L (ed), 2005 | 1 | A particular focus of the study was to assess the value of local-level wetland resource use by wetland communities, in one of the Zambezi's largest wetland complexes. Most of the population in the Barotse Floodplain depend on a mixed livelihood strategy. |
| Emerton, and Bos, 2004 | 1 | This study used avertive expenditure techniques to value the flood attenuation services of Muthurajawela Marsh, a coastal peat bog. |
| Emerton and Muramira, 1999 | 1 | 1950s: establishment of a Controlled Hunting Area<br>1960s and 1970s: establishment of a Game Reserve<br>1980s: establishment of a National Park<br>1990s: community wildlife conservation in Lake Mburo National Park |
| Emerton, et al., 1998 | 1 | Although existing pressure on biological resources is generally low in Djibouti, the demands of a rapidly expanding population and diversifying economy may in the future make biodiversity exploitation increasingly unsustainable. |
| Gerrard, P., 2004 | 2(1) | The goal of the study is to demonstrate the importance of sustainable management of wetland areas; Methods included price based approaches used to measure both direct and indirect use values. A large portion of the wetland has been converted to rice cultivation. |
| Karanja, et al., 2001 | 2 | The Pallisa District has been identified as having important wetlands, which face severe threats, and requires urgent management interventions. Most of these wetlands have been drained on a large scale to pave way for developments. |
| Kasthala, et al., 2008 | 2 | The goal of the project is that pro-poor approaches to the conservation and sustainable use of threatened wetlands are strengthened through improved capacity, awareness and information on the biodiversity and livelihood value of aquatic ecosystems. |
| Kumari, K., 1996 | 2 | The peat swamps would be protected and not drained. …consequently the peat swamps are not in an entirely satisfactory ecological condition. There are indications of a worrying decline in the overall integrity and status of the wetland habitats. |
| Lant, and Roberts, 1990 | 1 | CV is used to estimate the recreational and intrinsic benefits of improved river quality in selected river basins… |
| Loth, P. (ed) 2004 | 1 | It was anticipated that re-inundation would have the following positive impacts on floodplain goods and services… |
| Ly, et al., 2006 | 1 | This report presents the results of an economic study of the willingness-to-pay (WTP) of recreational visitors to the Djoudj National Bird Park (DNBP),… |
| Mallawaarachchi, et al., 2001 | 1 | The available area of wetlands has declined faster than the area of tea-tree woodlands, and this relative scarcity value is reflected in the estimates. (quantity reduced and quality unchanged) |
| Meyerhoff, and Dehnhardt, 2004 | 1 | Our study about the river Elbe shows that riparian wetlands provide significant benefits that should be considered in river basin management decisions. |
| Mmopelwa, et al., 2009 | 2 | While the Okavango Delta has sustained local inhabitants through ecosystem goods and services for centuries, the functioning of the Delta's natural ecosystem has been threatened over the past two decades by several water "development" initiatives. |
| Phillips, A. (ed), 1998 | 2 | A \$0.2 option value per hectare per annum figure for savannah and wetland systems, is used to calculate Uganda's annual option value for the pharmaceutical industry. |
| Rosales, et al., 2005 | 1 | Sekong is part of the Central Annamites, which has been identified as one of five priority regions of the WWF's Ecoregion Conservation Program in Indochina. |
| Schuijt, K., 2002 | 2 | The economic and social impacts of large dams on African communities living on the floodplains have been mostly adverse; The underlying cause of much wetland degradation is information failures. |
| Thibodeau, and Ostro, 1981 | 1 | This study presents a structure for analyzing the economics of wetland preservation and applies it to the Charles River watershed. The benefit cost method is applied to the wetland for wetland preservation. |
| Tong, et al., 2007 | 2 | The results showed that the potential value at the Sanyang wetland was 55,332 yuan ha$^{-1}$ yr$^{-1}$, while the current value was only 5807 yuan ha$^{-1}$ yr$^{-1}$. |
| Turpie, et al., 1999 | 2 | People there traditionally used resources according to their customary laws based on indigenous knowledge systems. Now, few incentives exist for communities to be involved in natural resource management… |
| Turpie, J.K., 2000 | 2 | The area is rich in wildlife and plant resources, However, there is concern that the area's biodiversity is under threat from unsustainable use of these resources … |

Our rules to determine the quality state are given as follows. Firstly, if there is direct description of wetland degradation, then assign quality code 2 to the wetland site. Otherwise, code 1. Secondly, if the mentioned human activities are obviously able to lead to wetland degradation, then assign code 2. Thirdly, if the market price method is used to measure values, then it increases our confidence in assigning a code 2 when the degradation occurs. Finally, if the values are estimated for an assumed ideal state of the wetland, then the quality should be assigned 1. This often happens when wetland restoration projects are evaluated with the benefit-cost method at the wetlands which face the threat of loss or degradation. Proper application of these rules is subject to the understanding of the meaning of the values in the articles.

### 4.3    Meta-analysis considering wetland quality

In general, a meta-analytic benefit function can be written as:

$$y_i = a + b_S X_{Si} + b_E X_{Ei} + b_C X_{Ci} + \mu_i$$

where $yi$ measures the benefit of ecosystem site $i$ and is an dependent variable. Independent variables or explanatory variables include:

$X_{Si}$, characteristics of valuation studies,

$X_{Ei}$, characteristics of ecosystem and

$X_{Ci}$, characteristics of socio-economic and geographical context;

$b_S$, $b_E$ and $b_C$ are the vectors containing the coefficients of the explanatory variables;

$a$ is a constant.

In this analysis, $X_E$ contain wetland properties like wetland size, type and services, which are information that can be retrieved from the TEEB database, and wetland quality, which is gained through tracing the original literature as mentioned before. $X_S$ mainly refer to the methods used for economic valuation, while $X_{Ci}$ include gross national income (GNI) per capita and national population density. Information is available in the TEEB database.

Dummy variables are essentially a device to classify data into mutually exclusive categories (Gujarati, 2003). Dummy variables apply in this analysis to distinguish the values produced by different wetland types, ecosystem services types and wetlands with different qualities, and to reveal the difference due to different valuation methods adopted. Table 4 gives an overview of the explanatory variables and data points available for the analysis.

**Table 4**   Explanatory variables used in the meta-regression model

| Group | Variable | Type | Levels / measurement unit | N |
|---|---|---|---|---|
| Study ($X_S$) | Valuation method | Nominal | Avoided Cost | 8 |
| | | | Contingent Valuation | 5 |
| | | | Direct market pricing | 41 |
| | | | Factor Income / Production Function | 5 |
| | | | Mitigation and Restoration Cost | 3 |
| | | | Replacement Cost | 7 |
| | | | Travel Cost | 1 |
| Wetland ($X_E$) | Wetland size | Ratio | in Hectares | 70 |
| | Wetland Quality | Binary | 0/1 | 70 |
| | Wetland type | Nominal | Floodplains | 23 |
| | | | Peat-wetlands | 6 |
| | | | Swamps / marshes | 17 |
| | | | Wetlands [unspecified] | 24 |
| | Service provided | | Climate | 1 |
| | | | Extreme events | 6 |
| | | | Food | 18 |
| | | | Genepool | 4 |
| | | | Medical | 2 |
| | | | Ornamental | 1 |
| | | | Raw materials | 18 |
| | | | Recreation | 5 |
| | | | Soil fertility | 1 |
| | | | Waste | 8 |
| | | | Water | 5 |
| | | | Water flows | 1 |
| Socio-Economic Context ($X_C$) | GNI per capita | Ratio | 2007 US\$ person$^{-1}$ year$^{-1}$ in Nation in which wetland is found | 70 |
| | Population density | | Population per sq km in Nation in which wetland is found | 70 |

N = number of observations for each variable or variable level

The model fit was significantly improved by transforming the dependent variable and the explanatory variables of wetland size, GNI per capita and population density into natural logarithms. Table 5 presents the results, which were obtained using this logarithmic model. In this model, the coefficients of the explanatory variables expressed in natural logarithms represent the percentage change in the dependent variable, given a one percent change in the explanatory variable (Gujarati, 2003), while the coefficients of the dummy variables correspond to the difference of the log values among different categories of each characteristic group.

**Table 5** Results obtained with the meta-regression model of wetland values

|  | Variable | Coefficient | p-value |
|---|---|---|---|
|  | (Constant) | 1.645 | .374 |
| Study ($X_S$) | Avoided Cost | 5.182** | .001 |
|  | Contingent Valuation | 2.201 | .176 |
|  | Direct market pricing | .140 | .899 |
|  | Replacement Cost | 2.944** | .020 |
| Wetland ($X_E$) | Size(ln) | -.183* | .060 |
|  | Quality | 1.562** | .038 |
|  | Floodplains | .670** | .020 |
|  | Peat-wetlands | 1.199* | .078 |
|  | Swamps / marshes | .799 | .965 |
|  | Climate | -2.989 | .242 |
|  | Extreme events | -1.944 | .188 |
|  | Food | 1.324 | .237 |
|  | Genepool | -1.598 | .334 |
|  | Medical | -2.609 | .160 |
|  | Raw materials | . 388 | .741 |
|  | Recreation | -1.339 | .293 |
|  | Soil fertility | -2.454 | .275 |
| Socio-Economic Context ($X_C$) | GNI per capita (ln) | .257 | .189 |

a. Dependent Variable: LogY2007
b. OLS results. $R^2 = .676$; *Adj. $R^2 = .561$. F = 5.905; p-value = 0.000.* Significance is indicated with ***, ** and * for 1, 5 and 10% statistical significance levels respectively.

Comparing the coefficients of different wetland services, it can be found that the values for the service of food tend to be the highest among all the ecosystem services. Among different wetland types, the values provided by marshes and flood plains are at the middle. For the valuation methods, the coefficients collectively indicate that the direct market pricing method tends to give a lower value than most of other methods. Finally, the coefficient of wetland size shows a significant negative relationship between the values and the wetland size, while another explanatory expressed as logarithms tells that the values are positive related to the GNI per capita, even if it is not statistically significant.

The coefficient of the wetland quality is 1.562, indicating the difference of the log values between those wetlands functioning naturally and those that have been degraded due to human activities. Considering of how quality codes are achieved for this analysis, we could also interpret this as a considerable difference between wetland potential and current degraded wetland values. P value for this positive coefficient is 0.038, showing the difference is statistically significant at the 5% level.

# 5    Conclusions

It is not perplexing that under similar socio-economic and geographical conditions, those wetlands which function naturally generate higher value services than those that have been degraded due to human activities. However, this simple but critical feature of wetlands had not been embodied in most of the comprehensive statistical analyses of wetland valuation studies. This research implemented the integration of the wetland quality into the meta-regression analysis, where the relative but globally consistent quality information applies.

However, there remain uncertainties. Firstly, the perceived quality tends to lack precision and moreover, there is no distinct quality information for each function where services are based. Secondly, the effect of the wetland quality on values might be influenced by other factors, which are not taken in the analysis. Some examples are sustainability levels of wetland use in the primary studies, scarcity of urban wetlands and productivity of the wetlands due to their geographical conditions.

Accurate ecosystem valuation can contribute much to the improvement of policy design by enabling comparative analysis of intervention practices. In order to achieve accurate estimates through a meta-function benefit transfer, as an alternative to primary studies for non-valued wetlands, it is required to find a more valid and accurate meta function to reflect the essential relationship between values and determinants. This research has made effort in this way by exploring the effect of wetland quality on economic value of ecosystem services. Much improvement could be realised when further primary valuation studies, especially those complemented with wetland quality information, are made available.

**Reference**

Adamus, PR, Clairain EJ, Smith RD and Young RE, 1987. Wetland Evaluation Technique (WET); Volume II: Methodology. Operational Draft Technical Report Y-87; U.S. Army Engineers Waterways Experiment Station, Vicksburg, MS.

Arrow KJ, Dasgupta P and Mäler KG, 2004. Are we consuming too much? J. Econ. Perspective, 18, 147–172.

Barbier EB, 1993. Sustainable Use of Wetlands: Valuing Tropical Wetland Benefits—Economic Methodologies and Applications. The Geographical Journal, 159, 22–32.

Barbier E, Acreman MC and Knowler D, 1997. Economic Valuation of Wetlands: A Guide for Policy Makers and Planners. Gland, Switzerland: Ramsar Convention Bureau.

Bateman IJ and Jones AP, 2003. Contrasting Conventional with Multi-Level Modeling Approaches to Meta-Analysis: Expectation Consistency in U.K. Woodland Recreation Values. Land Economics, University of Wisconsin Press, 79(2), 235-258.

van den Bergh, JCJM., Button, KJ, 1997. Meta-analysis of environmental issues in regional, urban and transport economics. Urban Studies, 34 (5-6), 927–944

Brander L, Florax RJGM and Vermaat JE, 2006 The Empirics of Wetland Valuation: A Comprehensive Summary and a Meta-Analysis of the Literature. Environmental and Resource Economics, 33, 223-250.

Brander LM, Van Beukering PJH and Cesar HJS, 2007. The recreational value of coral reefs: a meta-analysis. Ecological Economics, 63, 209–218.

Brander L, Brauer I, Gerdes H, Ghermandi A, Kuik O, markandya A, Navrud S, Nunes P, Schaafsma M, Vos H, Wagtendonk A, 2009. Scaling up cosystem service value: using GIS and meta- analysis for value transfer, IVM and EEA, Copenhagen (Report).

Brinson, MM, Smith RD, Whigham DF, Lee LC, Rheinhardt RD, and Nutter WL, 1998. Progress in development of the hydrogeomorphic approach for assessing the functioning of wetlands. Paper read at Proceedings from the INTECOL International Wetland Conference, at Perth, Australia.

Brouwer R, Langford IH, Bateman IJ and Turner RK, 1999. A meta-analysis of wetland contingent valuation studies, Regional environmental change, 1(1), 47-57.

Costanza R, d'Arge R, de Groot R, Farber S, Grasso M, Hannon B, Limburg K, Naeem S, O'Neill RV, Paruelo J, Raskin RG, Sutton P, van den Belt M, 1997. The value of the world's ecosystem services and natural capital. *Nature*, 387, 253-260.

Cole CA, 2006. HGM and wetland functional assessment: Six degrees of separation from the data? *Ecological Indicators*, 6(3), 485-493.

Dasgupta P, 2010. Nature's role in sustaining economic development, Phil. Trans. R. Soc. 365, 5-11.

Dasgupta P and Mäler G, 2000. Net national product, wealth, and social well-being. Environ. Dev. Econ, 5, 69–93.

Emerton L (ed), 2005. Values and Rewards: Counting and Capturing Ecosystem Water Services for Sustainable Development. IUCN Water, Nature and Economics Technical Paper No. 1, IUCN - The World Conservation Union, Ecosystems and Livelihoods Group Asia.

Eshet T, Baron MG and Shechter M, 2007. Exploring benefit transfer: disamenities of waste transfer stations. *Environmental and Resource Economics*, 37, 521-47.

Ghermandi A, van den Bergh JCJM, Brander LM, de Groot HLF, and Nunes PALD, 2010. Values of natural and human-made wetlands: A meta-analysis. *Water Resour. Res.*, *46*, W12516, doi:10.1029/2010WR009071.

Gujarati D, 2003. Basic Econometrics, 4th Edition. McGraw-Hill/Irwin.

Glass GV, 1976. Primary, secondary, and meta-analysis of research. Educational Researcher, 5, 3-8.

Heimlich RE, Weibe KD, Claassen R, Gadsy D, House RM, 1998. Wetlands and agriculture: private interests and public benefits, Resource Economics Division, E.R.S., USDA, Agricultural Economic Report 765.10.

Karr JR, Fausch KD, Angermeier PL, Yant PR, and Schlosser IJ, 1986. Assessing biological integrity in running waters a method and its rationale, Illinois Natural History Survey Special Publication 5, Champaign, IL.

Kumar P and Chen H, 2014. Estimating the Welfare Loss of Climate Change Impact on Corals, in Handbook on the Economics of Ecosystem Services and Biodiversity, Edward Elgar, pp 93-112

Ledoux L, Turner RK, Mathieu L and Crooks S, 2001. Valuing ocean and coastal resources: practical examples and issues for further action. Paper presented at the Global Conference on Oceans and Coasts at Rio+10: Assessing Progress, Addressing Continuing and New Challenges, UNESCO, Paris, 3-7 December 2001.

Lindhjem HL and Navrud HL, 2008. How reliable are meta-analyses for international benefit transfers? Ecol Econ 66, 425–435.

Lodge TE, Hillestad HO, Carney SW, and Darling RB, 1995. Wetland quality index (WQI): a method for determining compensatory mitigation requirements for ecologically impacted wetlands. Proceedings of the American Society of Civil Engineers South Florida Section

Mäler KG, Anyar S and Jansson A, 2009. Accounting for ecosystems. environment and resource economics, 42, 39-51.

Maltby, E. 1986. Waterlogged Wealth, Why Waste The Worlds Wet Places; Earthscan: London, UK.Maltby E (ed), 2009. Functional Assessment of Wetlands: Towards Evaluation of Ecosystem Services. Woodhead Publishing, Cambridge.

Maltby 2022. The Wetlands Paradigm Shift in Response to Changing Societal Priorities: A Reflective Review. Land 11(9):1526

Millennium Ecosystem Assessment (MA), 2005. Findings from Responses Working Group. Island Press: Washington DC.

Miller SJ and Wardrop DH, 2006. Adapting the floristic assessment index to indicate quality anthropogenic disturbance in central Pennsylvania wetlands. Ecological Indicators, 6(2), 313-326.

Navrud S and Ready R (Eds), 2007. Environmental Value Transfer: Issues and Methods, Springer.

Nelson, J., Kennedy, P., 2009. The use (and abuse) of meta-analysis in environmental and natural resource economics: an assessment. Environ. Resour. Econ. 42 (3), 345–377.

Pearce DW and Atkinson G, 1993. Capital theory and the measurement of sustainable development: an indicator of weak sustainability, Ecological Economics, 8, 103-108.

Prasad SN, Ramachandran TV, Ahalya N, Sengupta T, Kumar A, Tiwari AK, Vijayan VS and Vijayan L, 2002. Conservation of wetlands in India - a review, Tropical ecology, 43(1), 173-186.

Rosenberger RS and Phipps TT, 2007. Correspondence and convergence in benefit transfer accuracy: A meta-analytic review of the literature. In S. Navrud and R.Ready (eds.), *Environmental Values Transfer: Issues and Methods.* Dordrecht, The Netherlands: Kluwer Academic Publishers.

Smith PGR and Theberge JB, 1987. Evaluating natural areas using multiple criteria: theory and practice. *Environ. Manage*,11(4), 447-460.

Smith RD, 1994. Hydrogeomorphic approach to assessing wetland functions developed under Corps' research program. Vicksburg, MS.

Smith RD, Amman A, Bartoldus C and Brinson MM, 1995. An approach for assessing wetland functions based on hydrogeomorphic classification, reference wetlands, and functional indices. WRP-DE-9. Vicksburg, MS: U.S. Army Engineer Waterways Experiment Station.

Smith VK and Kaoru Y 1990. Signals or Noise? Explaining the Variation in Recreation Benefit Estimates, American Journal of Agricultural Economics (May), 419–433

Smith VK and Pattanayak S, 2002. Is Meta-Analysis a Noah's Ark for Non-Market Valuation? Environmental & Resource Economics, 22(1), 271-296.tiglitz J, Amartya sen and Fitoussi JP (eds), 2009. Report by the Commission on the Measurement of Economic Performance and Social Progress, Paris. (www.stiglitz-sen-fitoussi.fr)

Seilheimer TS and Chow-Fraser P, 2006. Application of the  wetland fish index to northern great lakes marshes with emphasis on georgian bay coastal wetlands. Can. J. Fish. Aquat. Sci, 63, 354-366.

TEEB, 2008. The Economics of Ecosystems and Biodiversity, Policy Updates, 2009, European Union, Brussels. (www.teebweb.org)

Turner KG, et al., 2016. A review of methods, data, and models to assess changes in the value of ecosystem services from land degradation and restoration. Ecol. Model., 319, 190–207.

Turner R, van den Bergh J, Söderqvist T, Barendregt A,  van der Straaten J, Maltby E and van Ierland E, 2000. Ecological-economic analysis of wetlands: scientific integration for management and policy. Ecological Economics, 35(1), 7-23.

Verhoeven JT, Setter TL. Agricultural use of wetlands: opportunities and limitations. Ann Bot. 2010 Jan;105(1):155-63. doi: 10.1093/aob/mcp172. PMID: 19700447; PMCID: PMC2794053.

Wright T, Tomlinson J, Schueler T, Capiella K, Kitchel A and Hirschman D, 2006. Article 1 of the wetlands & watersheds article series: Direct and indirect impacts of urbanization on wetland quality. Prepared for Office of Wetlands, Oceans, and Watersheds, U.S. Environmental Protection Agency. Ellicott City, Md.: Center for Watershed Protection. Available at:  http://www.northinlet.sc.edu/training/media/resources/Article1Impact%20Urbanization%20Wetland%20Quality.pdf

Zedler JB. and Kercher S, 2005. Wetland Resources: Status, trends, ecosystem services and restorability, Annual Review of Environment and Resources, 30, 39-74.